\documentstyle[a4,12pt]{article}
\begin{document}
\parskip=5pt plus 1pt minus 1pt

\begin{flushright}
{\bf DPNU-97-15}\\
{March 1997}
\end{flushright}

\vspace{0.2cm}

\begin{center}
{\large\bf Is the Cabibbo-Kobayashi-Maskawa Matrix Symmetric} \\
{\large\bf at the GUT Scale?}
\end{center}

\vspace{0.3cm}

\begin{center}
{\bf Zhi-zhong Xing} \footnote{Electronic address: xing@eken.phys.nagoya-u.ac.jp}
\end{center}
\begin{center}
{\it Department of Physics, Nagoya University, Chikusa-ku, Nagoya 464-01, Japan}
\end{center}

\vspace{3cm}

\begin{abstract}
By use of the one-loop renormalization group equations and current experimental
data, we study the off-diagonal asymmetries of the Cabibbo-Kobayashi-Maskawa (CKM) matrix
at the GUT scale in the framework of the standard model as well as its two Higgs
and supersymmetric extensions. It is concluded that the possibility of a
symmetric CKM matrix at the GUT scale has almost been ruled out.
\end{abstract}

\vspace{2cm}

\begin{center}
({\it Accepted for publication in J. Phys. G})
\end{center}

\newpage

In the standard $\rm SU(3)\times SU(2)\times U(1)$ gauge model, quark mixing
and $CP$ violation can be naturally described by the $3\times 3$ 
Cabibbo-Kobayashi-Maskawa (CKM) matrix $V$ \cite{CKM}. The geometrical structure of
$V$ is characterized by its two off-diagonal asymmetries, one about the
$V_{ud}-V_{cs}-V_{tb}$ axis (denoted by $\Delta_1$) and the other about the 
$V_{ub}-V_{cs}-V_{td}$ axis (denoted by $\Delta_2$) \cite{Xing1}:
\begin{eqnarray}
\Delta_1 & \equiv & |V_{us}|^2 - |V_{cd}|^2 \; = \; |V_{cb}|^2 - |V_{ts}|^2 
\; = \; |V_{td}|^2 - |V_{ub}|^2 \; , \nonumber \\
\Delta_2 & \equiv & |V_{us}|^2 - |V_{cb}|^2 \; = \; |V_{cd}|^2 - |V_{ts}|^2 
\; = \; |V_{tb}|^2 - |V_{ud}|^2 \; .
\end{eqnarray}
Some speculation has recently been made on the possibility of $\Delta_1 =0$
and its consequences at low-energy scales \cite{Xing1,Tanimoto,Kobayashi}.
This possibility is of particular interest on the point that it
allows $V$ to be a symmetric matrix (about the $V_{ud}-V_{cs}-V_{tb}$ 
axis), which can be fully described in terms of only three independent 
parameters. A careful analysis shows that $\Delta_1 \sim 10^{-5} - 10^{-4}$ 
and $\Delta_2 \geq 400 \Delta_1$ are favoured by current experimental 
data \cite{Xing2}.
This implies that the possibility of a symmmetric CKM matrix has almost been 
ruled out at low-energy scales.

\vspace{0.3cm}

A natural question to be asked is if there still exists the possibility
for $\Delta_1 =0$ at a superheavy scale, e.g., the grand unification
theory (GUT) scale. In a specific GUT framework $\Delta_1 =0$ is 
possible to hold as the consequence of a hidden symmetry, and the
nonvanishing value of $\Delta_1$ at low-energy scales may come purely from 
the different evolution effects of $|V_{us}|^2$ and $|V_{cd}|^2$
(or other relevant CKM matrix elements).
Although many renormalization-group analyses of the CKM matrix were
made in the literature (see, e.g., \cite{Ma,Babu}), a careful
look at the evolution behaviour of $\Delta_1$ and $\Delta_2$ has been lacking.

\vspace{0.3cm}

A symmetric CKM matrix at (or above) the GUT scale should be theoretically
intertesting. Since the CKM matrix is basically determined by quark mass
matrices, its nine elements are expected to be calculable in terms of 
the quark mass ratios (as well as the possible $CP$-violating phases) in
the framework of a theory more fundamental than the standard model. At
low-energy scales we have known that it is impossible to derive the exact
symmetric CKM matrix from any quark mass ansatz, due to the observed 
difference between quark mass ratios $m_u/m_c$ and $m_d/m_s$ as well as
that between $m_c/m_t$ and $m_s/m_b$ \cite{Xing3}. 
Imposing specific but reasonable
symmetries on quark mass eigenvalues at a superheavy scale (e.g., the
left-right symmetry together with a $SU(2)_R$ symmetry of Yukawa couplings
and a $U(1)$ horizontal family symmetry \cite{Ross}), one would
have some chance to obtain the symmetric quark flavor mixing matrix.
Reversely, if a symmetric CKM matrix at the GUT scale were favored by 
current experimental data, then it could provide us some hints towards
the dynamical details or symmetries of quark mass generation.

\vspace{0.3cm}

In this note we shall investigate the magnitudes of $\Delta_1$ and
$\Delta_2$ at the GUT scale by use of the one-loop renormalization
group equations and current data. For $\Delta_1 \sim 10^{-5} - 10^{-4}$
at low-energy scales, we find that the possibility of $\Delta_1=0$ 
has almost been ruled out at $\mu = 10^{16}$ GeV in the framework of 
the minimal supersymmetric standard model (MSSM) or at $\mu =10^{14}$ GeV
in the framework of the two Higgs doublet model (2HM) for all perturbatively 
allowed values of $\tan\beta$. In the context of the standard model (SM),
$\Delta_1$ increases with energy while $\Delta_2$ decreases with energy.
The evolution effect of $\Delta_2$ is only at the percent level for every
model under discussion.

\vspace{0.3cm}

Without loss of generality, the one-loop renormalization group equations for 
the CKM matrix elements can be written as \cite{Babu}:
\begin{eqnarray}
16 \pi^2 \frac{{\rm d}}{{\rm d} t} |V_{i\alpha}|^2 & = & 3 c
\left [ \sum_{j\neq i} \sum_{\beta = d,s,b} \frac{f^2_i + f^2_j}
{f^2_i - f^2_j} ~ f^2_{\beta} ~ {\rm Re} \left (V_{i\beta} V_{j\beta}^* 
V_{j\alpha} V_{i\alpha}^* \right ) \right . \nonumber \\
&  & \left . + ~ 
\sum_{\beta\neq \alpha} \sum_{j = u,c,t} \frac{f^2_{\alpha} + f^2_{\beta}}
{f^2_{\alpha} - f^2_{\beta}} ~ f^2_j ~ {\rm Re} \left (V_{j\beta}^* V_{j\alpha}
V_{i\beta} V_{i\alpha}^* \right ) \right ] \; , 
\end{eqnarray}
where $t \equiv \ln (\mu / M_Z)$ with $M_Z = 91.187$ GeV,
the Latin (Greek) indices stand for the up (down) quarks, $c$ is a 
model-dependent coefficient, $f_i$ ($i=u,c,t$) and $f_{\alpha}$
($\alpha = d,s,b$) are the eigenvalues of the Yukawa coupling matrices.
The evolution equations of $f_i$, $f_{\alpha}$ and $f_{\alpha^{\prime}}$ 
($\alpha^{\prime} = e,\mu, \tau$)
read \cite{Babu}:
\begin{eqnarray}
16 \pi^2 \frac{{\rm d} f_i}{{\rm d} t} & = & f_i \left [ 3 \sum_{j=u,c,t}
f^2_j  +  3 a \sum_{\beta = d,s,b} f^2_{\beta}  +  a \sum_{\beta^{\prime}
= e,\mu,\tau} f^2_{\beta^{\prime}}  -  G_U  +  \frac{3b}{2} f^2_i  +  
\frac{3c}{2} \sum_{\beta = d,s,b} f^2_{\beta} |V_{i\beta}|^2 \right ] \; ,
\nonumber \\
16 \pi^2 \frac{{\rm d} f_{\alpha}}{{\rm d} t} & = & f_{\alpha} \left [ 
3 a \sum_{j=u,c,t} f^2_j  +  3 \sum_{\beta = d,s,b} f^2_{\beta}  +  
\sum_{\beta^{\prime} = e,\mu,\tau} f^2_{\beta^{\prime}}  -  G_D  +  
\frac{3b}{2} f^2_{\alpha}  +  \frac{3c}{2} \sum_{j = u,c,t} f^2_j 
|V_{j\alpha}|^2 \right ] \; , \nonumber \\
16 \pi^2 \frac{{\rm d} f_{\alpha^{\prime}}}{{\rm d} t} & = & 
f_{\alpha^{\prime}} \left [ 3 a \sum_{j=u,c,t} f^2_j  + 
3 \sum_{\beta = d,s,b} f^2_{\beta}  +  \sum_{\beta^{\prime} = e,\mu,\tau} 
f^2_{\beta^{\prime}}  -  G_E  +  \frac{3b}{2} f^2_{\alpha^{\prime}}
\right ] \; ,
\end{eqnarray}
where $a$, $b$ and $c$ are model-dependent coefficients, and the
quantities $G_F$ ($F=U,D,E$) are functions of the gauge couplings
$g_n$ ($n=1,2,3$):
\begin{equation}
G_F \; = \; \sum_{n=1}^{3} \left ( C^F_n g^2_n \right ) \; , ~~~~~~~~~~
8 \pi^2 \frac{{\rm d} g^2_n}{{\rm d} t} \; = \; b_n g^4_n \; .
\end{equation}
Here $C^F_n$ and $b_n$ are also model-dependent coefficients. The values of
all such coefficients have been listed in Table 1 for the SM, 2HM and MSSM,
respectively. 

\vspace{0.3cm}

Table 1: The values of the model-dependent coefficients in the one-loop
renormalization group equations for the CKM matrix elements, the eigenvalues
of the Yukawa coupling matrices and the gauge couplings.

\begin{center}
\begin{tabular}{c|ccc} \hline\hline 
Coefficients	& SM	& 2HM	& MSSM \\ \hline  \\
$\left \{ a ~,~ b ~,~ c \right \}$ 	
& $\{ 1 ~,~ 1 ~,~ -1 \}$ 
& $\left \{ 0 ~,~ 1 ~,~ \frac{1}{3} \right \}$ 
& $\left \{ 0 ~,~ 2 ~,~ \frac{2}{3} \right \}$ \\ \\
$\left \{ b_1 ~,~ b_2 ~,~ b_3 \right \}$ 	
& $\left \{ \frac{41}{6} ~,~ -\frac{19}{6} ~,~ -7 \right \}$
& $ \{ 7 ~,~ -3 ~,~ -7 \}$
& $ \{ 11 ~,~ 1 ~,~ -3 \}$ \\ \\
$\left \{ C^U_1 ~,~ C^U_2 ~,~ C^U_3 \right \}$ 	
& $\left \{ \frac{17}{12} ~,~ \frac{9}{4} ~,~ 8 \right \}$ 
& $\left \{ \frac{17}{12} ~,~ \frac{9}{4} ~,~ 8 \right \}$ 
& $\left \{ \frac{13}{9} ~,~ 3 ~,~ \frac{16}{3} \right \}$ \\ \\
$\left \{ C^D_1 ~,~ C^D_2 ~,~ C^D_3 \right \}$ 	
& $\left \{ \frac{5}{12} ~,~ \frac{9}{4} ~,~ 8 \right \}$ 
& $\left \{ \frac{5}{12} ~,~ \frac{9}{4} ~,~ 8 \right \}$ 
& $\left \{ \frac{7}{9} ~,~ 3 ~,~ \frac{16}{3} \right \}$ \\ \\
$\left \{ C^E_1 ~,~ C^E_2 ~,~ C^E_3 \right \}$ 	
& $\left \{ \frac{15}{4} ~,~ \frac{9}{4} ~,~ 0 \right \}$ 
& $\left \{ \frac{15}{4} ~,~ \frac{9}{4} ~,~ 0 \right \}$ 
& $\left \{ 3 ~,~ 3 ~,~ 0 \right \}$ \\ \\ \hline\hline
\end{tabular}
\end{center}

\vspace{0.3cm}

Next we carry out a numerical analysis of the evolution effects for 
$\Delta_1$ and $\Delta_2$ by use of the above equations. The initial
values of gauge couplings at $\mu =M_Z$ (i.e., $t=0$) are taken as
$g^2_1 = 0.127$, $g^2_2 = 0.42$ and $g^2_3 = 1.44$ \cite{Babu}. We
take the light quark masses at $\mu = 1$ GeV to be $m_u = 5.6$ MeV,
$m_d = 9.9$ MeV, $m_s = 199$ MeV, $m_c = 1.35$ GeV and $m_b = 5.3$
GeV \cite{Gasser,PDG}. These masses evolve up to $\mu = M_Z$ due only to
gauge interactions, and this running effect can be approximately 
described by a common factor 0.58 for their values at $\mu = 1$ GeV
\cite{Masip}. The top-quark mass is typically taken as $m_t = 180$
GeV at $\mu = M_Z$. The charged lepton masses can be found from
ref. \cite{PDG}. The eigenvalues of the Yukawa coupling matrices are 
the ratios of the fermion masses to the Higgs vacuum expectation value $v$ 
(normalized to 175 GeV), and they may depend on the ratio of the
two vacuum expectation values (defined by $\tan\beta$) in the 2HM
and MSSM. We also choose $|V_{ud}| =
0.9744$, $|V_{us}|=0.2205$ and $|V_{cb}|=0.040$ as input parameters
\cite{PDG}. Since $\Delta_1 \sim 10^{-5} - 10^{-4}$ is allowed at
low-energy scales \cite{Xing1,Xing2}, we typically take 
$\Delta_1 = 1\times 10^{-5}$,
$5\times 10^{-5}$ and $1\times 10^{-4}$ at $\mu = M_Z$ in our 
calculations. The main results about the magnitudes of $\Delta_1$
and $\Delta_2$ at the GUT scales are illustrated in figs. 1 $-$ 4.
Some discussions are in order.

\vspace{0.3cm}

(1) One can see from fig. 1 that the value of $\Delta_1$ increases 
with energy in the context of the SM. Thus there is no possibility
for $\Delta^{\rm sm}_1$ to be vanishing at any superheavy scale. 

\vspace{0.3cm}

(2) In the framework of the 2HM, the magnitude of $\Delta_1$ decreases
with energy, and its behaviour as a function of $\tan\beta$ at the GUT 
scale $\mu =10^{14}$ GeV is shown in fig. 2. We observe that 
$\Delta^{\rm 2hm}_1$ remains to be nonvanishing for the low-energy inputs
$1\times 10^{-5} \leq \Delta_1 \leq 10^{-4}$. If the value of $\Delta_1$ 
at $\mu = M_Z$ were much smaller than $10^{-5}$, $\Delta_1^{\rm 2hm}=0$ 
would be possible in the perturbative region of $\tan\beta$.

\vspace{0.3cm}

(3) It is more interesting to look at the evolution effect of $\Delta_1$
in the framework of the MSSM. At the GUT scale $\mu = 10^{16}$ GeV,
the changes of $\Delta^{\rm mssm}_1$ with $\tan\beta$ are illustrated
in fig. 3. Clearly $\Delta_1^{\rm mssm} < \Delta_1^{\rm 2hm}$ for
the same input of $\tan\beta$. Taking $\Delta_1 =0$ at low-energy
scales, we find that $\Delta_1^{\rm mssm} < 0$ in the perturbative region 
of $\tan\beta$. This implies that there would exist the possibility
for $\Delta_1^{\rm mssm}=0$ if the low-energy value of $\Delta_1$ were 
well below $10^{-5}$ (e.g., $\Delta_1 \sim 10^{-7}$ or smaller).

\vspace{0.3cm}

(4) As indicated by current experimental data \cite{Xing1,Xing2}, the chance 
for $\Delta_1$ to be larger than $10^{-4}$ or smaller than $10^{-5}$ at 
low-energy scales is very tiny. Although the possibility of 
$|V_{us}| = |V_{cd}|$ has not been absolutely ruled out by direct
measurements, it is expected to be impossible from the combined data on
quark mixing and $CP$ violation \cite{PDG}. In this sense, we may conclude 
that the possibility of a symmetric CKM matrix at the GUT scales is almost
ruled out by current low-energy data.

\vspace{0.3cm}

(5) The evolution behaviour of $\Delta_2$ with energy is opposite to that of 
$\Delta_1$ in the context of the SM, 2HM or MSSM. Since $\Delta_2 \approx
|V_{us}|^2$ at low-energy scales, it has no possibility to be vanishing
at any GUT scale. For illustration, we plot the ratio $\Delta_2 (\mu)
/ \Delta_2 (M_Z)$ as a function of $\tan\beta$ in fig. 4, where 
$\Delta_2 (M_Z) = 0.047$ has been used as the input. We see that the
running factor of $\Delta_2$ in each model is only at the percent level.

\vspace{0.3cm}

In summary, we have studied the off-diagonal asymmetries of the CKM
matrix ($\Delta_1$ and $\Delta_2$) at the GUT scale by use of the one-loop 
renormalization group equations. For $\Delta_1 \sim 10^{-5} - 10^{-4}$
at low-energy scales, the possibility of $\Delta^{\rm mssm}_1=0$ (or
$\Delta_1^{\rm 2hm}=0$) has almost been ruled out at $\mu = 10^{16}$
GeV (or $\mu = 10^{14}$ GeV) for all perturbatively allowed values of
$\tan\beta$. In contrast, $\Delta_1$ increases with energy in the framework
of the SM. The magnitude of $\Delta_2$ decreases with energy in the context 
of the SM, but it increases with energy in the context of the 2HM or the MSSM.

\end{document}